\documentclass[conference]{IEEEtran}
\usepackage{cite}
\usepackage{amsmath,amssymb,amsfonts}
\usepackage{algorithmic}
\usepackage{graphicx}
\usepackage{textcomp}
\usepackage{xcolor}
\def\BibTeX{{\rm B\kern-.05em{\sc i\kern-.025em b}\kern-.08em
    T\kern-.1667em\lower.7ex\hbox{E}\kern-.125emX}}
\begin{document}

\title{The Generation of Software Security Scoring Systems Leveraging Human Expert Opinion\\
}

\author{\IEEEauthorblockN{Peter Mell}
\IEEEauthorblockA{\textit{Computer Security Division} \\
\textit{National Institute of Standards and Technology}\\
Gaithersburg, United States \\
https://orcid.org/0000-0003-2938-897X}
}

\maketitle

\begin{abstract}
While the existence of many security elements in software can be measured (e.g., vulnerabilities, security controls, or privacy controls), it is challenging to measure their relative security impact. In the physical world we can often measure the impact of individual elements to a system. However, in cyber security we often lack ground truth (i.e., the ability to directly measure significance). 
In this work we propose to solve this by leveraging human expert opinion to provide ground truth. Experts are iteratively asked to compare pairs of security elements to determine their relative significance. On the back end our knowledge encoding tool performs a form of binary insertion sort on a set of security elements using each expert as an oracle for the element comparisons. The tool not only sorts the elements (note that equality may be permitted), but it also records the strength or degree of each relationship. The output is a directed acyclic `constraint' graph that provides a total ordering among the sets of equivalent elements. Multiple constraint graphs are then unified together to form a single graph that is used to generate a scoring or prioritization system. 

For our empirical study, we apply this domain-agnostic measurement approach to generate scoring/prioritization systems in the areas of vulnerability scoring, privacy control prioritization, and cyber security control evaluation.
\end{abstract}

\begin{IEEEkeywords}
software, measurement, security, scoring, prioritization
\end{IEEEkeywords}

\section{Introduction}
\label{sec.introduction}

The existence of many software security properties can be measured. For example, we can measure the presence of vulnerabilities, security controls, and privacy controls on a system. But it is challenging to measure the significance of any one element compared to another; ad hoc approaches have been developed (e.g., \cite{CVSSv3.1}, \cite{OWASP}, \cite{CWETop25}, \cite{galhardo2020measurements}, \cite{chew2008performance}, \cite{liu2011vrss}, \cite{liu2012improving}, \cite{wang2009security}, \cite{wang2011improved}, and \cite{CWRAF}), but it is not usually possible to demonstrate that any particular methodology is correct (or even useful on occasion). However, in many hard sciences there is ground truth (measurements obtained through observation) to enable such comparisons. For example, in an electrical circuit two unlabeled resistors could be compared by measuring their effects on voltage. One could then empirically determine an ordering among a set of unlabeled resistors. Even better, a scoring system could be developed that would label each resistor with the exact resistance it provided. How can this be done for a set of security properties to order or score them as to their significance? How do we measure their relative 'security' to each other?

Unfortunately, often what prevents us is a lack of ground truth (the ability to directly measure the significance of a security element). One possible solution path is to tie security elements to expected dollar losses or probability of intrusion/vulnerability (e.g., \cite{DoDCAR} and \cite{zhang2011empirical}), but in general such data is not available. Yet rating the security significance of a discovered vulnerability or security control that needs to be implemented is an operational necessity; these determinations are made every day on networks using ad hoc approaches. What is needed is some form of ground truth that will enable the measurement of the relative security between security properties.

In this work, we answer such questions through leveraging human experts. Experts are iteratively asked to compare pairs of security elements to determine their relative significance (e.g., less than, much greater than, or equal to). They are asked O($n$log$n$) questions where $n$ is the number of security elements to be compared. On the back end our knowledge encoding tool performs a form of binary insertion sort on a set of security elements using the human as an oracle for each element comparison. The tool not only sorts the elements (note that equality may be permitted), but it also records the strength or degree of each relationship (e.g., greater than or much greater than). The relationships are recorded in a directed acyclic graph where each node represents an element, and the edges represent the degree of the recorded relationship between two elements. Edges of degree 0 are used to group equal elements; this enables us to place each element in an equivalency set (possibly of size 1). We call these output graphs `constraint graphs'. The constraint graphs produced by the knowledge encoding tool will provide a total ordering of the associated equivalency sets.

Multiple experts may each use the knowledge encoding tool multiple times to produce a set of constraint graphs. We measure the differences between constraint graphs to 1) ensure that each expert provides consistent input and 2) detect outlier constraint graphs. Sets of constraint graphs are then unified through a voting procedure. The voting works by evaluating the defined relationship between all pairs of security elements in each input graph. The pairs are ordered based on how strongly the experts agreed with respect to the relationship between any two elements. Note that often experts will not have explicitly compared a pair of elements; thus, we rely on the transitivity of the total ordering of the equivalency sets to determine the relationship from the constraint graph. The output is a single unified constraint graph (that is directed and acyclic but that may have lost its total ordering of the equivalency sets due to disagreements among the experts).

Finally, we use the unified constraint graph to generate either a prioritization or scoring system (depending upon the security domain being evaluated). The generation of prioritization systems is trivial provided that the total ordering was maintained in generating the unified graph (if not, that must be resolved). The generation of a scoring system is more involved. First, the human operator must choose minimum required distances between generated scores given the differing degrees between the security elements specified by the unified graph. Most valid settings will allow for an infinite number of `rational' scoring systems to be constructed. We say that a scoring system is rational if it does not violate any edge in the associated unified constraint graph. Next, the human operator may optionally `peg' certain elements to specific values. After that, our scoring generation tool suggests a rational scoring system using a straightforward greedy algorithm that takes into account the unified constraint graph as well as any pegged values. Further customization can be done by the human editor pegging additional values (within the constraints defined) and then recomputing suggested scores for the remaining unpegged security elements.

For the empirical portion of our research, we applied our approach to two security domains. We used the Forum of Incident Response and Security Teams' (FIRST) Common Vulnerability Scoring System (CVSS) vectors to generate an alternate scoring system \cite{CVSS}. We also used the National Institute of Standards and Technology (NIST) Privacy Framework (PF) to generate a prioritization of privacy controls \cite{boeckl2020nist}. We also implemented an interface to analyze the NIST Cyber Security Framework (CSF) but did not do a significant study in this domain \cite{barrett2020approaches}.

The remainder is of this paper is organized as follows. Section \ref{sec.design} presents the design of the system. Section \ref{sec.experiments} presents our two empirical studies using the CVSS and PF. Section \ref{sec.limitations} then highlights the limitations of this approach, and Section \ref{sec.futurework} explains planned future work. Section \ref{sec.conclusion} concludes.


\section{Scoring Generator System Design}
\label{sec.design}

Our scoring generator system is composed of three tools: the knowledge encoding tool, the constraint graph unification tool, and the scoring generation and prioritization tool. Each tool feeds its output into the next tool and thus each tool represents a stage in the scoring generation process.


\subsection{Knowledge Encoding Tool}
\label{sec.design.encoding}

The knowledge encoding tool is a domain-agnostic tool for representing human knowledge as directed acyclic graphs; in particular it represents degrees of significance between a set of elements within a knowledge domain. In our experiments, the elements are taken from a chosen security domain; our tool currently has modules to represent the CVSS, PF, and CSF. The tool iteratively presents a human expert with pairs of security elements to be compared. The expert then designates which is more severe, significant, or important (depending upon the security domain being analyzed). Optionally, the tool may be configured to allow the expert to express the degree of inequality (e.g., less than or much less than) and/or to express equality. After answering a series of questions that include all the domain elements at least once, the tool outputs a representation of the expert's knowledge.

On the back end, the answers provided by the expert are used to generate a directed acyclic graph where the nodes represent individual security elements and the edges represent the relationships between elements. The edges represent `greater than' relationships (or equality if permitted by the tool configuration). For the `greater than' relationships, the degree of minimum distance may be represented (e.g., greater than or much greater than). These degrees are represented using the natural numbers (with 0 representing equality). 

The algorithm used to construct the graphs is a modified binary insertion sort. Unprocessed elements are iteratively added to the graph using a binary insertion sort algorithm (leveraging an O(log n) binary search to find the correct location for the new element). However, the human expert is used to calculate the comparisons instead of the computer. An edge is added to the output graph after every comparison, showing the degree of the `greater than' relationship (or equality if that is permitted). 

Newly introduced elements that are designated as equal to an already processed element are added as children to the already processed element (using an edge of degree 0). No further comparisons are done using these child nodes; the parent represents this particular `equivalency set' of elements. Thus, each equivalency set is a collection of nodes defined by the human as equally significant, having a single parent with zero or more children. When identifying a node within an equivalency set to use for a comparison we considered taking a random node, but this could cause an unintentional shift in interpretation of the equivalency set by the human. This is because even though the elements are marked as equal they may not actually be exactly equal.  

The last important point is that when adding an unprocessed element, the first comparison made is done against a random equivalency set. This is done to more evenly distribute edges throughout the graph; otherwise the binary insertion sort, always starting with the middle node, generates graphs where the edges mostly bifurcate the set of nodes into two highly connected clusters.

The produced graph is guaranteed by the use of the binary insertion sort to have a longest path that includes all nodes that are parents of equal children or that have no children. As mentioned previously, each of these nodes represent an `equivalency set' (set of elements that are marked as equal), possibly of size 1. The outputted directed acyclic graph then totally orders the equivalency sets. Note however that it does more than totally order, because it may provide the minimum degree of distance between each equivalency set.

An example graph is shown in Figure \ref{fig:CVSS-Scores-9-3}. The black nodes on the longest path represent a set of elements that are totally ordered. The green edges represent `greater than' relationships and the black edges represent `much greater than' relationships. The yellow child nodes are those marked by the expert as being equal to their parents. This parent to child equivalency is represented by a very light blue edge.

\begin{figure}
\centerline{\includegraphics[scale=.6]{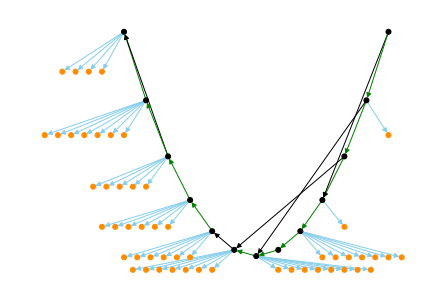}}
\caption{Example Output Graph (for Section \ref{sec.experiments.CVSS}, this is a Unified CVSS Graph of 9 Constraint Graphs from 3 Experts)}
\label{fig:CVSS-Unified-9-3}
\end{figure}


\subsection{Constraint Graph Unification Tool}
\label{sec.design.unification}

The constraint graph unification tool merges multiple constraint graphs into a single directed acyclic graph. This can be used to unify the knowledge representations of multiple experts. It can also be used to better represent the knowledge of a single expert by the expert evaluating a security domain multiple times and then unifying their resultant constraint graphs.

Note that all graphs to be unified must have the same set of security elements. Thus, they have the same number of nodes but likely a different set of edges (including a different number of edges) due to our use of our modified human-directed binary insertion sort graph generation algorithm. However, since they are guaranteed by construction to be directed and acyclic, every input graph provides the relationship between every pair of nodes (using edge transitivity). The nodes are totally ordered (while optionally allowing for equality).

The unification algorithm starts by enumerating all pairs of nodes. Each two nodes $x$ and $y$ could form the pairs ($x$,$y$) and ($y$,$x$), but only one will be chosen (which one is arbitrary). For each chosen pair, there will be a directed path in each input graph between the parent nodes representing the equivalency sets in which $x$ and $y$ reside. If parent($x$) is equal to parent($y$), then $x$ and $y$ are equal because they reside in the same equivalency set. If there is a path from parent($x$) to $y$, then the relationship is `greater than' (we do not use the degree of the edges until the final scoring generation). Likewise, if the path is from parent($y$) to $x$ then the relationship is `less than'. For each input graph and for each chosen pair $x$ and $y$, we record the number of times we observe `less than', equal, and `greater than'. The pair ($x$,$y$) is labeled with these measurements, which represent human expert votes for each relationship.

Next we adjust the votes for each pair as follows. For each vote for both `less than' and `greater than', we subtract 1 vote from both `less than' and `greater than' and we increase the votes for equal by 1. Thus, two votes that disagree on the direction of the relationship get consolidated into one vote for equal. The intuition is that if two experts can't agree on the direction of a relationship, then the two elements are more likely to be equal.

Each pair is then labeled with the maximum number of votes for any particular relationship and all the pairs are ordered by this maximum value from highest to lowest. This ordering is the priority that will be used to process the pairs and add them to the output graph. 
The idea is that if a pair receives a large number of votes for any particular relationship, that pair should receive priority over other pairs (with fewer votes) in being added to the constructed output graph.

The pairs are then processed in priority order and the nodes of each pair and the designated maximum relationship are added to the, initially empty, output graph. If a pair to be added has two relationships tied for maximum (equal combined with either `less than' or `greater than'), this this pair is not added to the output graph; it is marked as disputed. If the addition of a pair would cause the output graph to have cycles (thus creating a contradiction with previously added higher priority and higher confidence pairs), it is not added; it is marked as contradictory. In this way we process all $(n^2)/2-n$ chosen pairs, adding to the output graph first those with higher votes (representing more confidence). 

One final complexity is that the output graph generation algorithm will change the specific edges but preserve the encoded logic. It does this to simplify the output graph. All nodes that are equal will be constructed as children of one particular parent node. All edges that would have connected to the children are changed to connect to the parent (since the parent represents that equivalency set).

The result is an output graph that unifies the input constraint graphs. By construction, it will be directed and acyclic. However, unlike the input graphs it is not guaranteed to totally order the nodes. This is because experts may disagree, and the resulting votes may not provide enough information to totally order the nodes. The visual artifact of this loss of total ordering is that there will not exist a single longest path that includes the parent node for each equivalency set. Fortunately, this will not cause an issue for our scoring generation algorithm. For our prioritization algorithm, however, we had to develop heuristics to address this possibility.


\subsection{Scoring Generation and Prioritization Tool}
\label{sec.design.generation}

The scoring generation and prioritization tool takes as input a single unified constraint graph and outputs either a scoring system or a prioritization of the inputted security elements.

\subsubsection{Scoring Generation}
As discussed previously, the knowledge encoding tool may require experts to specify the degree of distance between pairs of security elements (e.g., specifying `much less than', `less than', equal, `greater than', and `much greater than'). These get translated into edges that are labeled with degrees of distance (e.g., 0, 1, and 2 for equal, `greater than', and `much greater than' respectively). There are no negative distances as, for example, `less than' simply maps to degree 1 with the edge having a reversed direction.

To generate scores for a particular input graph, a human operator must then specify the minimum scoring distance required for each degree of distance. They also must specify bounds on the minimum and maximum scores to be produced. Given a set of minimum distance values and bounds, there may be no valid solutions, one solution, or an infinite number of solutions. If the scoring requirements limit the number of decimal places used in produced scores, there may be a finite number of valid solution sets instead of an infinite number. A valid solution set is one that is rational (i.e., each node in the constraint graph can be labeled with the score of its security element without violating any edge constraints). 

\begin{figure}
\centerline{\includegraphics[scale=.6]{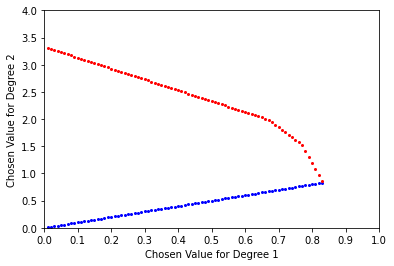}}
\caption{Valid Degree 2 Values (Minimum Score=0, Maximum Score=10) for Graph in Figure \ref{fig:CVSS-Unified-9-3}}
\label{fig:Degree2ValueChoices}
\end{figure}

We wrote a tool to calculate the set of valid minimum distance values that may be chosen. It works by evaluating the different combinations of edges that can be used to traverse a constraint graph. For graphs with edges of both degrees 1 and 2, it iterates over the possible degree 1 values and then calculates the associated maximum degree 2 value (the minimum degree 2 value is simply equal to the chosen degree 1 value). An example output is shown in Figure \ref{fig:Degree2ValueChoices}. The x-axis provides the possible degree 1 values that may be chosen (the minimum scoring distance for `greater than'). The top red line represents the maximum degree 2 value that may be chosen (the minimum scoring distance for `much greater than') and the lower blue line represents the minimum. Note that choosing a degree 2 value on the red line will result in exactly 1 scoring solution set being available, since the maximum range of scoring values will be needed in order to satisfy the edges in the constraint graph. This can be implemented in O($n^2$) time.

Next, an algorithm is executed on the input constraint graph that determines the minimum and maximum scoring assignments possible for each node. The outputted score for each security element is the mean of the maximum and minimum scores determined for each node. Minimal scores can be assigned by finding nodes whose parents have all been scored and assigning their score minimally based on the minimum distances specified by the parent to child edges. If a node has no parents, then it gets assigned the minimum valid score (usually 0). If a node has some parents that have been scored and others that have not, it waits to be processed until all its parents have been scored. For a node with a single parent, the node's score is the parent's score plus the minimum distance required by the edge degree (discussed previously). For a node with multiple parents, the node's score is calculated individually relative to each parent (as if the node had only one parent). Then the maximum value calculated is used for the node's minimum score. The node's maximal scores can be calculated with a slightly more complicated variant; both can be implemented in O($n^2$) time. 

Usually more than one rational scoring system can be generated; our algorithm outputs one of them. We do not claim that the outputted solution is necessarily superior to the others. Our algorithm spreads the scores out within the range of possible scores (as defined by the operator inputted minimum and maximum scores). This results in the less significant equivalency sets having scores assigned close to their minimums, while the more significant equivalency sets have scores assigned closer to their maximums.

Another important note is that the human operator may choose to peg certain nodes to certain values. Our algorithm can take this pegging into account when generating scores. Once a scoring system is generated, the operator may change a score within the defined range of maximum and minimum values for the corresponding node. This pegs the score within the system and the algorithm is then re-run to generate a new proposed scoring system. In this way a human operator can customize the scores and, through that customization capability, generate any of the possible rational solutions.

\begin{figure}
\centerline{\includegraphics[scale=.6]{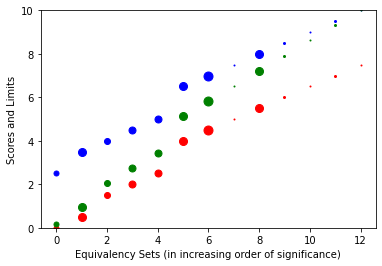}}
\caption{Generated CVSS Scores and Limits from the Unified Graph}
\label{fig:CVSS-Scores-9-3}
\end{figure}

An example scoring output graph along with the minimal and maximal values are shown in Figure \ref{fig:CVSS-Scores-9-3}. The x-axis values represent the different equivalency sets defined in the constraint graph (13 in this case). The size of each dot represents the size of the corresponding equivalency set (the number of security elements defined to have the same significance). The top blue nodes represent the maximal possible values. The bottom red nodes represent the minimal possible values. The green middle nodes represent the chosen values.

\subsubsection{Prioritization}
As discussed previously, the constraint graphs produced by the knowledge encoding tool will have a unique longest path that will contain exactly one node from every equivalency set. Thus, such graphs totally order the equivalency sets. To output a security element prioritization, we simply traverse the unique longest path outputting each of the security element associated with each visited node (along with the children of each node). No special handling needs to be done to output each equivalency set as each equivalency set is represented as a directed star subgraph with a single parent along the longest path. This means that all nodes equal to a node on the longest path are immediate children of that node.

On occasion a difficulty arises, because the unified constraint graphs are not guaranteed to be totally ordered. Looking ahead to our PF case study in Section \ref{sec.experiments.privacy}, Figure \ref{fig:PF-Unified-1-2} shows the unified constraint graph from 3 input graphs. The blue nodes not on the longest path of black nodes are those that could not be totally ordered given the information in the input graphs. We resolve this using the heuristic of assigning each non-totally ordered node as equivalent to the node on the longest path that is equidistant from the last node on the longest path (if it is a parent of a node on the longest path) or equidistant from the first node on the longest path (if it is a child of a node on the longest path). In our example, the rightmost blue node (which is a parent of a node on the longest path) gets assigned as equivalent with the initial node on the longest path (as both are distance 10 from the last node on the path). For the blue nodes on the left (which are children of nodes on the longest path), two of them are distance 9 and one distance 10 from the initial node on the longest path. They get assigned as equal to the nodes on the longest path at distance 9 and 10 respectively.


\section{Experiments}
\label{sec.experiments}

For our empirical study, we applied our methodology to two very different security domains. We discuss first our evaluation of the FIRST CVSS and then our evaluation of the NIST PF. In both cases we demonstrated the ability of our methodology to leverage human expert opinion to generate both a scoring system (in the case of the CVSS) and a prioritization system (in the case of the PF). 

We note that the purpose of this empirical study is to show how one can generate a scoring and prioritization system; the results are not intended to provide actual authoritative scoring and prioritization systems. Because of this, we did not formally define the necessary qualifications for the set of `human experts' for each security domain nor in all cases did we use actual human experts. Any real-world usage of this approach will need to both provide a rigorous definition and then ensure that all humans in the study comply with that definition.

\subsection{Common Vulnerability Scoring System}
\label{sec.experiments.CVSS}

The CVSS `provides a way to capture the principal characteristics of a vulnerability and produce a numerical score reflecting its severity' \cite{CVSS}. It is maintained by the FIRST, a global forum for incident response teams. The specification for CVSS version 3.1 is available at \cite{CVSSv3.1}. It defines 2496 vectors that can be used to describe vulnerabilities. Each vector consists of 8 metrics, each with from 2 to 4 possible metric values.

\begin{figure}
\centerline{\includegraphics[scale=.8]{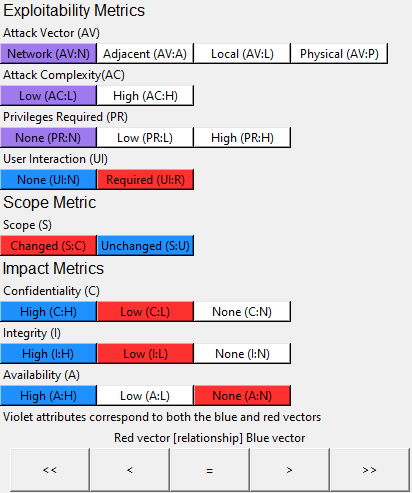}}
\caption{CVSS Analysis Visualization}
\label{fig:CVSS-Analysis-Console}
\end{figure}

Our user interface represents the CVSS vectors as shown in Figure \ref{fig:CVSS-Analysis-Console}. In this example, there are two vectors being presented for comparison by the human expert; the red vector is the new vector being inserted into the existing constraint graph and the blue vector is the current vector being compared against for the binary insertion sort. The red boxes represent metric values for the red vector; the blue boxes represent metric values for the blue vector; and the purple boxes represent metric values that apply to both the red and blue vector. The buttons at the bottom enable the expert to input the relationship between the red and blue vectors (much less than, less than, equal, greater than, or much greater than).

The author performed the analysis along with two volunteer researchers; while qualified scientists in the area of computer security, none of these 3 are experienced vulnerability analysts (although we continue to refer to them as `human experts' for consistency of terminology). Because of this, our empirical results will be used only to demonstrate the efficacy of the scoring generation system. The results provided are not intended to be compared against the CVSS scoring system or to suggest new scores.

Our team analyzed the top 65 frequently occurring CVSS version 3 vectors published within the National Vulnerability Database (NVD) \cite{NVD} as of 2020/07/08. This set of 65 vectors represented 90.03 \% of the published CVE vulnerabilities. Each expert performed the analysis 3 times (taking a mean of 1.4 hours to complete each session). 

Our goal in performing three sessions per expert was to determine 1) if the expert would provide consistent data and 2) if the system would produce consistent constraint graphs while presenting the expert different sets of comparisons\footnote{A comparison of the question sets asked between two runs with this dataset revealed only a 10 \% overlap of questions.}. We then used the data from all three experts to be able to compare the produced graphs and to have a dataset upon which to execute the voting algorithm to output a unified graph.

We analyzed each pair of graphs and calculated the number of inconsistent relationships (i.e., when they disagreed for a particular pair of vectors). With 65 vectors there were 2080 possible relationships ($(65^2-65)/2)=2080$). For pairs of graphs from the same expert, the mean number of inconsistent relationships was 440 (10.58 \%). This number reduced to 224 (5.37 \%) looking just at cases where the relationships point in opposing directions (i.e., one is less than and the other greater than). For pairs of graphs from different experts, the mean number of inconsistent relationships was 622 (14.96 \%). For opposing directions, it is 389 (9.34 \%). As expected, experts are more consistent with themselves than with other experts. However, even among different experts there was general agreement on the relationships defined.

We analyzed each pair of graphs and compared the produced CVSS vector orderings. Due to its construction, the equivalency sets in each constraint graph are guaranteed to be totally ordered. However, each graph has in general different nodes in each equivalency set. This creates a challenge in comparing outputted CVSS vector orderings, because the published algorithms for measuring differences in orderings require that both sets to be compared have exactly the same elements. We address this problem by converting the total ordering of equivalency sets from each graph into a total ordering of nodes (i.e., CVSS vectors). This entails fixing the ordering of vectors within each equivalency set (which doesn't cause problems, because they were all defined to be equal). We perform the arrangement of equivalent vectors to minimize the distances between the graphs. This is done by fixing the ordering of vectors within each equivalency set to match, as closely as possible, the total ordering of equivalency sets provided by the graph to be compared. For ordering comparisons, we use the Spearman algorithm \cite{Rankings}. This algorithm calculates the number of adjacent elements that would need to be swapped on an input ordered list in order to match a target list. For pairs of graphs from the same expert, the mean distance was 496. For pairs of graphs from different experts, the mean distance was 614. To provide context for these numbers, randomized lists of the same size produce a mean distance of 1432. These results show our approach is producing vector orderings between graphs that are much more consistent than random.

We imported each of the 3 generated graphs from the 3 experts (9 graphs total) into our constraint graph unification tool. These graphs had from 161 to 212 edges with a mean of 186; however when optimizing to reduce redundant edges, the range reduce to 107 from 178 edges with a mean of 138 (note that a redundant edge indicates an input from the expert that ultimately wasn't useful). They all had 65 nodes to represent the 65 CVSS vectors.

We then used our voting algorithm to unify the nine graphs into a single constraint graph. The 2080 pairs of vectors were compared and inputted into the voting scheme. As the voting scheme ran, it encountered 121 disputed relationships (where the voting was tied) and 219 contradictions (where a candidate with a lesser priority relationship violated the constraints already added to the graph by a higher priority relationship). These types of expected events are discussed in detail in Section \ref{sec.design.unification}. The unified graph is shown in Figure \ref{fig:CVSS-Unified-9-3}. This final graph has 65 nodes and 68 edges. 

While not guaranteed by the algorithm, it is interesting that the resulting constraint graph in Figure \ref{fig:CVSS-Unified-9-3} maintained the total ordering of the equivalency sets. This can be observed visually, because all nodes are either on a longest path or are children of a node on the longest path (using an edge type that marks them as equal to their parent).

\begin{figure}
\centerline{\includegraphics[scale=.6]{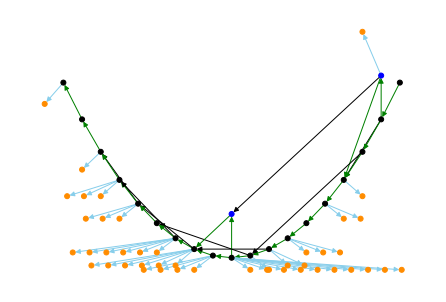}}
\caption{Unified CVSS Graph of 3 Constraint Graphs from 1 Expert Showing Non-Total Ordering of Equivalency Sets}
\label{fig:CVSS-Unified-3-1}
\end{figure}

We also used our voting algorithm independently on each set of 3 graphs from each expert. An example is shown in figure \ref{fig:CVSS-Unified-3-1}. Two of the three experts' unified graphs did not totally order the equivalency sets. This can be observed visually because there exists no longest path covering all equivalency sets. 

Lastly, we used our scoring generation algorithm to use the unified constraint graph to produce a rational scoring set (out of many valid solutions). Our input parameters specified a minimum 0.5 score difference for `greater than' relationships (the green arrows in figure \ref{fig:CVSS-Unified-9-3}) and 1.5 for `much greater than' (the black arrows in figure \ref{fig:CVSS-Unified-9-3}). These values were chosen somewhat arbitrarily, as many values produce rational solutions; they were chosen from the possible values shown in Figure \ref{fig:Degree2ValueChoices}. 

Decreasing the minimum distances enables more flexibility in adjusting the final scores; increasing them too much prohibits being able to produce a valid solution. The scores generated for the 13 equivalency sets are as follows: 0.6, 1.4, 2.3, 3.0, 3.7, 4.9, 5.6, 6.3, 7.0, 7.7, 8.6, 9.3, and 10.0. Note that the CVSS specification \cite{CVSSv3.1} requires scores to be between 0 and 10 with one decimal place (101 possible scores).

This output is shown visually in Figure \ref{fig:CVSS-Scores-9-3}. The size of each node represents the number of CVSS vectors within that equivalency set. The lower red nodes represent a minimum boundary (the lowest possible scores that could be generated) while the top blue nodes represent the maximum boundary. The green nodes represent the scores chosen by our algorithm. The equivalency sets are ordered in increasing significance; this left to right order matches the black nodes in figure \ref{fig:CVSS-Unified-9-3} also reading from left to right. Note the unusually large jump between equivalency sets 1 and 2 and then also between 4 and 5. These jumps represent the influence of the leftmost 2 black `much greater than' arrows in figure \ref{fig:CVSS-Unified-9-3}. The other 3 black arrows do not affect the scoring, because they span a sequence of `greater than' green arrows that collectively require a greater separation than the black arrow.



\subsection{Privacy Framework}
\label{sec.experiments.privacy}

\begin{figure*}
\centerline{\includegraphics[scale=.6]{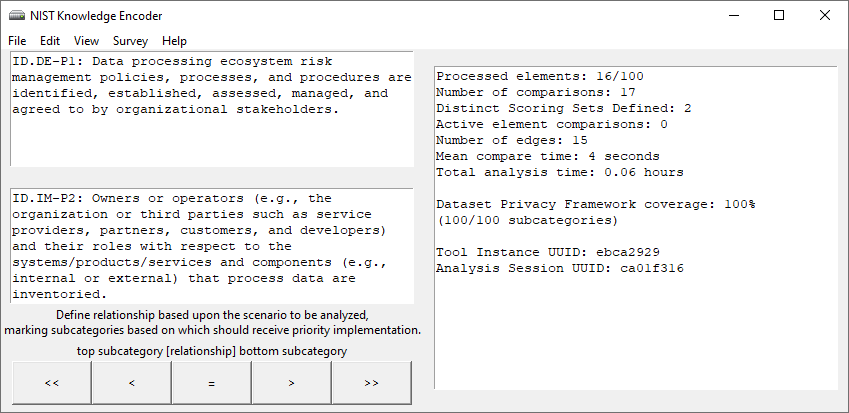}}
\caption{Privacy Framework Analysis Console}
\label{fig:PF-Console-Full}
\end{figure*}

The NIST PF is a tool `to help organizations identify and manage privacy risk to build innovative products and services while protecting individuals’ privacy' \cite{lefkovitz2020nist} \cite{boeckl2020nist}. It contains a hierarchy of controls at three levels of abstraction: functions, categories, and sub-categories. Our focus was to prioritize the 100 sub-categories for a particular privacy scenario. In doing this, we chose to limit human operators to inputting just $<$, $=$, and $>$ (two fewer options than with CVSS). 

Our volunteer privacy expert is a privacy professional. He constructed a scenario upon which to base his evaluation of the 100 privacy controls. The scenario was constructed to be specific enough to enable an evaluation of the controls but abstract enough to be useful to as broad a community as possible. The chosen scenario was the deployment of a COVID-19 digital exposure notification system within a company. When evaluating the controls, the expert assumed that no privacy controls had yet been implemented and thus was working to prioritize the 100 controls for implementation. Note that we did not use a scenario for CVSS, because CVSS inherently has its own scenario where it considers the impact of each vulnerability on the world at large. The console used by the expert to evaluate the 100 privacy controls is shown in Figure \ref{fig:PF-Console-Full}.  

\begin{figure}
\centerline{\includegraphics[scale=.6]{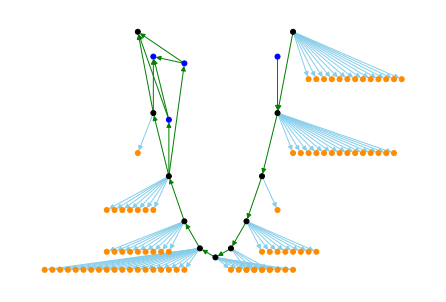}}
\caption{Unified Privacy Framework from 3 Constraint Graphs from 1 Expert}
\label{fig:PF-Unified-1-2}
\end{figure}

The expert analyzed the privacy sub-categories two times (taking a mean of 1.0 hour per session). The two graphs initially had a mean of 246 edges, but optimizations to remove redundant edges reduced that to a mean of 157 edges. There were 100 nodes to represent the 100 controls. The unified graph generated from the two input graphs is shown in Figure \ref{fig:PF-Unified-1-2}. As with Figure \ref{fig:CVSS-Unified-3-1}, it does not totally order the equivalency sets (as shown by the four blue nodes above the longest path). This only partial ordering may be due to us using our voting algorithm with just 2 input graphs; it is hard to use voting to break a tie with just 2 sets of votes.

Our goal with this experiment is to produce a prioritization list of privacy framework controls for implementation with the stated scenario. Thus, we did not generate scores. Instead, we used the unified constraint graph to generate equivalency sets of decreasing priority. As discussed previously, whenever a node is partially ordered (the ones in blue that are not on the longest path), we use the imperfect heuristic of including it in the set on the longest path equidistant from its parent or child. This yields 11 equivalency sets of privacy controls with decreasing priorities with the following set sizes: [14, 15+1, 2, 9, 5, 10, 20, 10, 8, 2+2, 1+1]. Note that the '+x' entries show our heuristic imperfectly adding the 4 partially ordered blue nodes to the totally ordered equivalency sets in order to be able to present a simple prioritization to the users. Had we access to our privacy expert to perform a 3rd analysis, we believe it likely that the output would be totally ordered, eliminating the need to use our imperfect heuristic.

The privacy expert spent 1 hour per analysis session (2 sessions in total) in generating the prioritization using our tool. Ideally we would have access to other experts, had them do this by hand, calculated the time, and analyzed the differences. Unfortunately, we do not have this data. What we can say is that the tool makes it easy for the human to perform the task (they just compare controls). More importantly though, it enables groups of human experts to work together on a prioritization (or scoring system) without having to coordinate and/or negotiate their analysis. 


\section{Limitations}
\label{sec.limitations}

Our scoring approach is domain agnostic and thus has wide applicability; however, the domain to be analyzed must be small enough that a human can analyze the vectors (i.e., perform $nlogn$ comparisons for the binary insertion sort). This effectively means that the inputs to the system must be nominal or ordinal measurements with a constrained set of possibilities. If the measurement is at the interval or ratio level, then it may be necessary to abstract and group ranges of values into categories related at the ordinal level of measurement. 

The NIST privacy framework contains only 100 controls (more precisely sub-categories), which makes this tractable. In the future, we'd like to analyze the NIST CSF sub-categories; there are only 98. For CVSS, each input vector contains a group of eight ordinal metrics that each have less than five possible values. This produces 2496 vectors, but we did not need to analyze all of them because the frequency distribution follows a power law type curve. Just 66 CVSS vectors covered 90 \% of the publicly published vulnerabilities in the NVD (as of 2021/01/08). 126 vectors covers 95 \%; 320 covers 99 \%. 



\section{Related and Future Work}
\label{sec.futurework}

We have been applying this approach to CVSS using human experts from the FIRST CVSS Special Interest Group (SIG). The SIG maintains the CVSS industry standard; we have two collaborative projects with them. The first is to calculate how much the CVSS v3.1 scoring system deviates from the closest rational scoring system generated from our approach using CVSS SIG experts. This work will be published late 2022 as NISTIR 8409 \cite{Mell2022}. The second collaborative project is to use our approach to assist the CVSS SIG in creating the scoring system for the upcoming CVSS version 4.0.

We also plan to investigate how one can map not yet analyzed security elements to a group of analyzed and scored elements. For example, there are several thousand CVSS vectors that our experts did not analyze; this was not a significant problem, because these vectors occur at extremely low frequencies (or not at all). However, we believe that through machine learning clustering techniques we can map the not yet analyzed vectors to the scored vectors with some measurable degree of accuracy.




\section{Acknowledgements}
\label{sec.relatedwork}

We would like to thank Loic Lesavre and Priam Varin from the NIST for their time spent analyzing CVSS vectors. In addition, we have great appreciation for Dylan Gilbert (also from NIST) who designed the privacy scenario and then performed the privacy control analysis.

\section{Conclusion}
\label{sec.conclusion}

In this work we have developed and demonstrated a new domain-agnostic technique for generating scoring systems. We have shown how human knowledge can be queried and encoded by asking an expert to iteratively compare pairs of elements from a particular domain. For the encoding, we modify the textbook binary insertion sort algorithm by 
\begin{itemize}
    \item using a human brain as the central processing unit (CPU) to power the algorithm,
    \item allow for equality and the recording of distance constraints,
    \item and adding initial comparison node randomization to more evenly spread the edges through the resultant knowledge graph.
\end{itemize}

The result is a directed acyclic graph that contains a unique longest path that totally orders the equivalency sets. Our constraint graph unification tool then combines multiple of these graphs and uses a voting algorithm to generate a unified constraint graph. Finally, our scoring generation and prioritization tool generates human customizable scoring systems consistent with the inputted constraint graph as well as prioritizations. 

We applied this domain-agnostic approach to the security domain and performed experiments using the CVSS, the PF, and (to a much lesser extent) the CSF.

In this work, we have demonstrated the usability and functionality of this approach. What we did not accomplish was to 1) prove the accuracy of the generated scoring systems / prioritizations and 2) perform a rigorous statistical study on the consistency and/or differences between multiple encoding of both a single expert and a group of experts. 

The former is unfortunately not generally possible, as no ground truth exists. We rest our defense of the approach in postulating that an expert will be able to accurately compare the relative significance of two elements. Empirically, we did find modest differences in the comparison results produced by the same expert when analyzing the same data multiple times. To reduce such errors, we suggest that each expert analyze each dataset three times.

The latter issue was not addressed in this work because, for the CVSS, the author did not have access to true domain experts and, for the PF, did not have enough domain experts (as well as the time of the single expert available) to perform any rigorous statistical studies.

Lastly, we note another major limitation of this work. This work only applies to knowledge domains that are small enough that a human can analyze the vectors. Mostly likely, the inputs need to be nominal or ordinal measurements with a constrained set of possibilities. To help mitigate this limitation, future work will explore combining this approach with machine learning where humans analyze the most prevalent elements and generate a scoring system for that subset; then machine learning approaches map the less frequent elements into the generated equivalency sets thereby enabling the scoring of all elements.

\end{document}